\documentclass[12pt,draftclsnofoot,onecolumn]{IEEEtran}

\usepackage{CJKutf8}
\usepackage{amsmath,amsfonts}
\usepackage{algorithmic}
\usepackage{algorithm}
\usepackage{array}
\usepackage{bm}
\usepackage{balance}
\usepackage{booktabs}
\usepackage{color}
\usepackage{diagbox}
\usepackage{setspace}
\usepackage{gensymb}
\usepackage{graphicx}
\graphicspath{{figures_LN/}}
\usepackage{siunitx}
\usepackage{stfloats}
\usepackage{caption}
\usepackage{textcomp}
\usepackage{threeparttable}
\usepackage{url}
\usepackage{verbatim}
\usepackage{makecell}
\usepackage{subfigure}
\usepackage{epstopdf}
\usepackage[letterpaper, top=2.5cm,bottom=3.5cm, left=3cm, right=2.5cm, marginparwidth=1.75cm]{geometry}
\makeatletter

\newcommand{\Rmnum}[1]{\expandafter\@slowromancap\romannumeral #1@}
\makeatother

\begin{document}

\begin{CJK}{UTF8}{gbsn}	
	\title{ Robust DOA Estimation Based on Dual Lawson Norm for RIS-Aided Wireless Communication Systems \vspace{-0.25em}}
	
	\author{ Canping Yu, Yingsong Li*, Liping Li, Zhixiang Huang, Qingqing Wu, Rodrigo C. de Lamare \vspace{-2em}

\thanks{This work was supported by National Natural Science Foundation of China (Grant No. 62071002). }

\thanks{Canping Yu, Yingsong Li, Liping Li and Zhixiang Huang are with the Key Laboratory of Intelligent Computing and Signal Processing Ministry of Education, Anhui University, Hefei 230601, Anhui, China (e-mail: liyingsong@ieee.org), \itshape(Corresponding author: Yingsong Li).}
  \thanks{Q. Wu is with the Department of Electronic Engineering, Shanghai Jiao Tong University, 200240, China (e-mail: qingqingwu@sjtu.edu.cn).}
  \thanks{Rodrigo. C. de Lamare is with the Centre for Telecommunications Research (CETUC), Pontifical Catholic University of Rio de Janeiro (PUC-Rio), G$\mathbf{\acute{a}}$vea, 22451-900, Brazil, and the School of Physics, Engineering and Technology, University of York, York, YO10 5DD, UK (e-mail: delamare@puc-rio.br).}}

\maketitle
\begin{abstract}
Reconfigurable intelligent surfaces (RIS) can actively perform beamforming and have become a crucial enabler for wireless systems in the future. The direction-of-arrival (DOA) estimates of RIS received signals can help design the reflection control matrix and improve communication quality. In this paper, we design a RIS-assisted system and propose a robust Lawson norm-based multiple-signal-classification (LN-MUSIC) DOA estimation algorithm for impulsive noise environments, which is divided into two parts: first, the non-convex Lawson norm is used as the error criterion along with a regularization constraint to formulate the optimization problem. Then, a Bregman distance based alternating-direction-method-of-multipliers (ADMM) is used to solve the problem and recover the desired signal. The second part is to use the multiple-signal-classification (MUSIC) to find out the DOAs of targets based on their sparsity in the spatial domain. In addition, we also propose a RIS control matrix optimization strategy that requires no channel state information (CSI), which effectively enhances the desired signals and improves the performance of the LN-MUSIC algorithm. A Cramér–Rao-lower-bound (CRLB) of the proposed DOA estimation algorithm is presented and verifies its feasibility. Simulated results show that the proposed robust DOA estimate algorithm based on the Lawson norm can effectively suppress the impact of large outliers caused by impulsive noise on the estimation results, outperforming existing methods.
\end{abstract}

	
	
    \section{Introduction}
The estimation of the direction-of-arrival (DOA) of signals is a fundamental task in the fields of communication systems and signal processing, as it has significant importance across diverse applications including wireless communication, radar systems, sonar systems, and microphone arrays.\cite{HH2018,GA2015}. In wireless communication systems, DOA estimation is employed to ascertain the direction in which the desired signal reaches the receiver~\cite{ZT2014}. By obtaining angular information about incoming signals, it is possible to direct the signal to the desired receiver and mitigate interference from other directions. This can significantly improve signal-to-noise-ratio (SNR) and consequently the system performance.

Recently, electromagnetic meta-material devices called reconfigurable intelligent surfaces (RIS) have attracted widespread attention. They can actively control the beams of reflected signals, alter the propagation conditions of communication channels and have become a crucial enabler for the wireless communication systems in the future \cite{RIS_future2021,RIS_future2022}. Integrating RIS into traditional communication systems can help to achieve NLOS communication and get simple system structure. In~\cite{SS_ANM2023}, a low-cost RIS-based passive sensing system was designed that requires only one antenna as a receiver for reliable DOA estimation. Furthermore, numerous studies have been constructed to implement the RIS-promoted DOA estimation. Reference~\cite{RISDOA2} designed a RIS-based DOA estimation scheme and presented a DOA estimation algorithm that combines the principles of maximum likelihood and compressed sensing. In~\cite{ULA2022}, a model for DOA estimation using a swarm of unmanned aerial vehicles (UAVs) equipped with RIS has been proposed. This model takes into account the positional disturbance of UAVs. Reference\cite{DNNRIS2023} puts forth a fresh methodology that merges deep neural networks (DNNs) with decoupled atomic norm minimization (ANM) to tackle the problem. This particular method was centered on addressing the challenge of estimating the 2-D DOA in scenarios involving RIS. Nevertheless, designing algorithms for DOA estimation, the majority of research studies made an assumption that the background noise follows a Gaussian-distribution. In actual communication systems, there are often non-Gaussian noises, such as impulsive noises. DOA estimation algorithms designed for Gaussian noise are often difficult to employ in a  stable way in the presence of impulsive noises.

Over the past few decades, there have been numerous reports on classic DOA estimation algorithms\cite{LC1997}, including subspace based methods \cite{music,Xin2012,Hui2011,jio_doa,alr_doa,mskaesprit,jidf,sjidf,jio,wljio,l1stap,locsme,okspme,lrcc,mcg,sbstap,rdstap,misc,misc2,dfsub}, maximum likelihood techniques \cite{CS2006}, and compressive sensing-based methods \cite{DL2006,dce,listomp}, all of which play important roles in various fields. Based on these classical DOA estimation methods, researchers have conducted many in-depth studies, such as reducing the computational complexity of algorithms, saving system resources\cite{shizi2023}, improving the accuracy of algorithms\cite{shijin2023}, and improving the robustness of algorithms in different scenarios\cite{shilu2023,shilu2}. In particular, several works have studied how to perform DOA estimation in impulsive noise environments\cite{DOAlu2009,DOAlu2013,DOAlu2017}, among which the norm minimization strategy based on compressive sensing is a very common method.

In the context of DOA-estimation algorithms based on compressive-sensing, we consider the received signal of a sensor array system as follows:
    \begin{equation}
    \begin{aligned}
    \mathbf{y} = \mathbf{A}\mathbf{x}+\mathbf{n},  
    \end{aligned}
    \end{equation}
where $\mathbf{y} \in \mathbb{C}^l$ is the received signal, $\mathbf{A} \in \mathbb{C}^{l \times m}$ denotes the {sensing matrix\cite{2007Compressive}}, $\mathbf{x} \in \mathbb{C}^m$ is the sparse signal, $\mathbf{n} \in \mathbb{C}^l$ is the background noise. The DOA estimation problem is typically modelled as a least-absolute-shrinkage-and-selection-operator (LASSO) model when $\mathbf{n}$ is a Gaussian noise vector according to:
    \begin{equation}
    \begin{aligned}
    {\min_{\mathbf{x}\in \mathbb{C}^m}} \|\mathbf{x}\|_1 + \rho \|\mathbf{A}\mathbf{x}-\mathbf{y}\|_2^2,
    \end{aligned}
    \end{equation}
where $\|\mathbf{x}\|_1$ denotes the $l_1$ norm, $\|\cdot\|_2$ is the $l_2$ norm, and $\rho$ is the regularization parameter.

The $l_2$ norm-based cost function can converge quickly with Gaussian background and is widely used, but its sensitivity to outliers makes it unsuitable for dealing with impulsive noises.  Many previous studies have used the $l_p$ norm as a measure of error to handle outliers in signals
    \begin{equation}
    \begin{aligned}
    {\min_{\mathbf{x}\in \mathbb{C}^m}} \|\mathbf{x}\|_1 + \rho \|\mathbf{A}\mathbf{x}-\mathbf{y}\|_p^p,
    \end{aligned}
    \end{equation}
if $p<1$, $l_p$ norm has a good suppression effect on outliers, but at this point, the cost function is not convex and smooth, it is often considered a non-deterministic polynomial (NP)-hard problem, which is difficult to solve.

In addition, many other robust fitting strategies have been proposed. In\cite{Lorentzian2023}, the use of the Lorentzian norm allows for the recovery of sparse signals from measurements that have been distorted by two types of impulsive noises. A weighted $l_2$ norm equivalent scheme has been reported in \cite{weightL22023}. The study in \cite{Hub_sinht2015} developed a Huber function to solve the multi-channel sparse recovery problem of complex measurements, which is robust against heavy-tailed non-Gaussian conditions. Recently, an adaptive filtering algorithm for robust echo cancellation using the Lawson norm was proposed in\cite{WS2020,Lawson2022}. The Lawson norm can approximate the $l_p$ norm, and unlike the non-smooth nature of the $l_p$ norm, it is continuously differentiable within the defined domain. The Lawson norm has not been investigated for DOA estimation problems so far. 

In this work, we consider a RIS-supported wireless system that exploits the ability of RIS to actively manipulate the amplitudes and phases of reflected signals to support NLOS communication between targets and antennas. In addition, RIS can save system resources and can perform DOA estimation of the target signals using only one receive channel snapshot. Given the disruptive nature of impulsive noises, we use the Lawson norm to construct the error criterion and the regularization constraint, mitigating the effect of outliers caused by impulsive noises. Subsequently, the constructed non-convex optimization problem is solved to recover the desired signal using Bregman's alternating-direction-method of multipliers (ADMM) strategy. Then, by exploiting target sparsity in the spatial domain, a Lawson-norm single snapshot multiple-signal-classification (LN-MUSIC) algorithm is developed to estimate the DOA of the signals. In addition, the RIS control matrix is designed by a random optimal state method (ROSM), which significantly reduces path loss impact on system performance. Finally, the simulations confirm the superior behavior of the proposed LN-MUSIC algorithm to existing techniques. The main contributions of this work are:

{1) We design a signal reconstruction algorithm based on the Lawson norm and propose a reliable solution within the framework of the ADM (Bregman's ADMM). This algorithm is capable of achieving reliable DOA estimation under impulsive noise interference without any requirement on the sparsity level of the signal. To the best of our knowledge, such an algorithm has not been presented in any prior research on DOA estimation or signal reconstruction.}

{2) During the iteration of the LN-MUSIC algorithm, the parameters of the Lawson criterion are adjusted and the optimal Bregman ADMM penalty parameter corresponding to the SNR is obtained using nonlinear fitting.}

{3) We have introduced an optimization scheme for the RIS control matrix, which does not need the Channel State Information (CSI) of the system, is straightforward to implement and can significantly enhance the performance of the LN-MUSIC algorithm. This approach has not been previously reported in the literature.}

4) The theoretical Cramér-Rao-lower-bound (CRLB) for DOA-estimation under impulsive noise is derived by leveraging the proposed sensor array system.

The paper is presented as follows: Section $\text{\Rmnum{2}}$ presents the DOA estimation model. Section $\text{\Rmnum{3}}$ describes the formulation of the optimization problem and corresponded solution procedure. Section $\text{\Rmnum{4}}$ discusses the CRLB for the DOA estimation with impulsive noises. Section $\text{\Rmnum{5}}$ gives a short discussion about the simulation results. The conclusion is given in Section $\text{\Rmnum{6}}$.

\textit{Notations}: Matrices are given by bold uppercase letters, while vectors are presented by bold lowercase letters, $(\cdot)^{\rm{T}}$ is transpose of matrices, $(\cdot)^{\rm{H}}$ denotes conjugate-transpose of matrices, $\text{diag}(\mathbf{a})$ is a diagonal of a matrix with vector $\mathbf{a}$ as dominant diagonal elements, $ \left \langle  \cdot \right \rangle $ is the vector inner product operator and $ \left \langle  \mathbf{a},\mathbf{b} \right \rangle = Re\left[\mathbf{b}^{\text{H}}\mathbf{a}\right] $, $Re\left[\cdot \right]$ is the real part of a complex number.

\section{System Model}

\begin{figure}[t]
    \centering
    \includegraphics[width=3.5in]{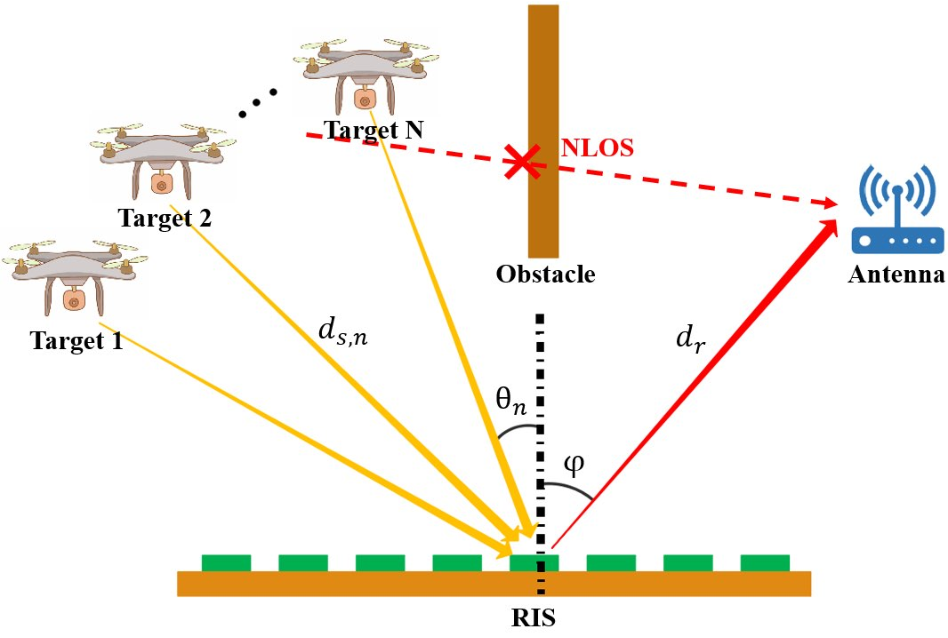}%
    \caption{The RIS-based wireless communication model.}
    \label{fig:model}
    \end{figure}

We considers a multiple-antenna wireless communication model \cite{mmimo,wence} with RIS assistance, which is discussed in Fig.\,\ref{fig:model}. There are $N$ targets, an antenna with only one channel propagation path to the receive antenna, and the direct path between them is blocked by an obstacle. The RIS comprises $M$ reflective elements arranged in a uniform linear array (ULA) structure. Its function is to reflect signals from targets to antenna, facilitating NLOS communication between them.
        
Based on the proposed model, consider the one-dimensional (1D) DOA estimation to obtain the DOA on the RIS. Defining the distance from the $n$-th $(n=0,1,\cdots,N-1)$ target to the RIS as $d_{s,n}$, the path transmission delay as $\tau_{s,n}$, and path-loss constant is $\alpha_n$. Let the signal from the $n$-th target be $s(t)$, then the superimposed signals gotten by the $m$-th element of the RIS {from the targets} is given by 
    \begin{equation}
    \begin{aligned}
    r_m \left ( t \right ) = \sum_{n=0}^{N - 1} \frac{\alpha_n}{d_{s,n}}s\left (t- \tau_{s,n}\right ) e^{j2 \pi m d \text{sin} \theta_n},
    \end{aligned}
    \end{equation}
where $\theta_n$ is DOA of the $n$-th target to the RIS and $d$ represents the wavelength normalised distance of the adjacent-elements on the RIS.

    \begin{figure}[t]
    \centering
    \includegraphics[width=3.5in]{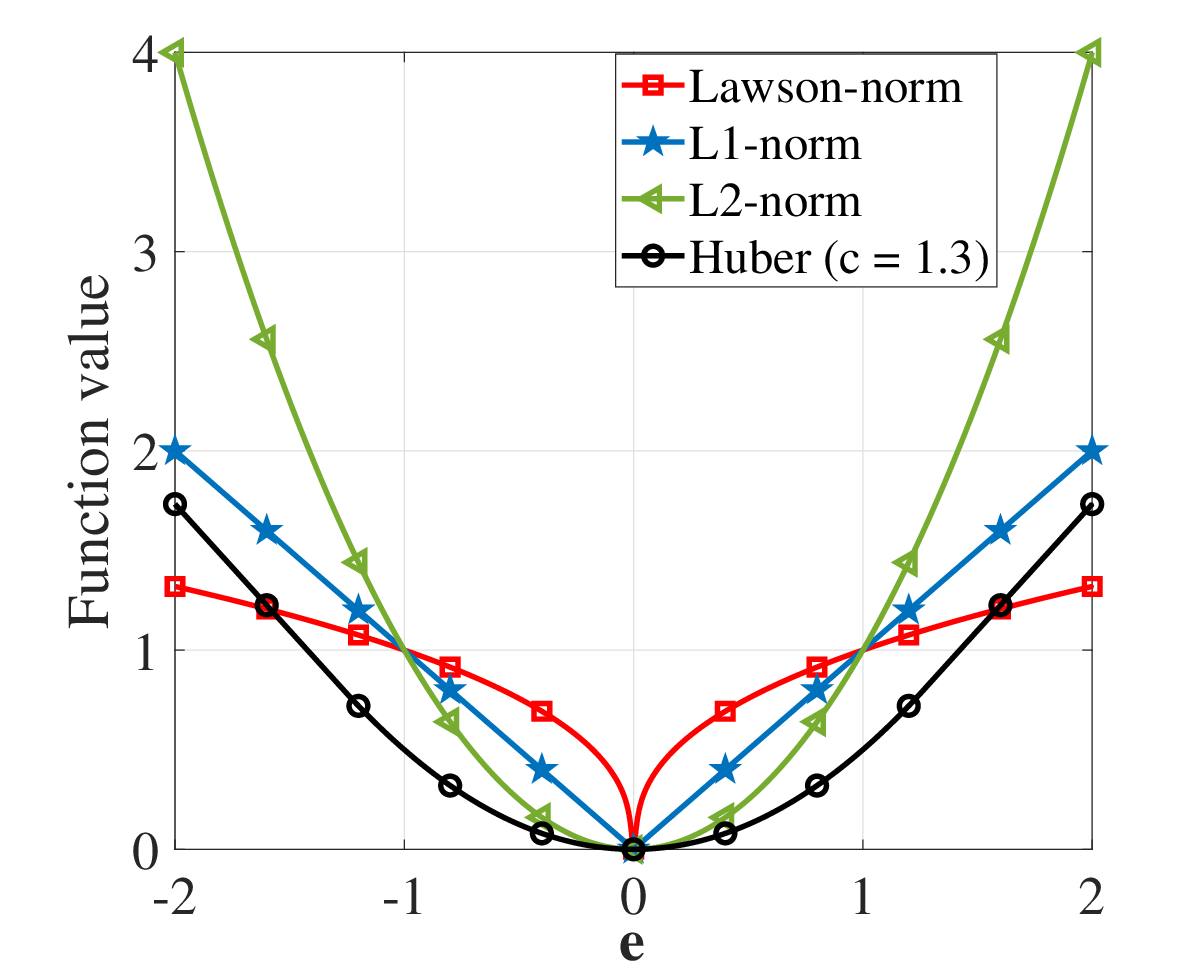}%
    \caption{Different norms as objective functions $(\lambda = 0.1)$.}
    \label{fig:norm_vs}
    \end{figure}
    
Each element within the RIS can have its own amplitude and phase settings, in order to achieve multiple measurements, enabling the state of each element in the RIS to be altered during $K$ different time slots (the time-slot period is $T$). Let the signal amplitude and phase change controlled by the $m$-th reflector unit during the $k$-th $(k=0,1,\cdots,K-1)$ time-slot be $A_{k,m}$ and $e^{j\phi_{k,m}}$ respectively. Defining the transmission delay between RIS and antenna as $\tau_r$, the distance as $d_r$, the path-loss constant as $\gamma$. RIS element's size is much smaller compared to signal bandwidth $B$, then the signal received by the antenna during the $k$-th slot is        
    \begin{equation}
    \begin{aligned}
    \label{eqn:ds_y}
    y_k \left ( t \right ) = \sum_{m=0}^{M-1} \sum_{n=0}^{N-1}\frac{\gamma}{d_r} A_{k,m} e^{j\phi_{k,m}} \frac{\alpha_n}{d_{s,n}}s\left (t - \tau_{s,n} -\tau_r \right ) e^{j2\pi m d \text{sin} \left (\theta_n + \varphi \right )}+v_k\left( t\right),
    \end{aligned}
    \end{equation}
where $\varphi$ is the direction between the RIS and the antenna and $v_k(t)$ is system noise for the $k$-th time slot. 
        
Suppose the state of each RIS element is changed at frequency $\frac{1}{T}$ $(BT \ll 1)$, the gotten signal data from the antenna in $K$ different time slots can be represented as a column vector of length $K$. Let us then define the following:        
    \begin{equation}
    \begin{aligned}
    \mathbf{y}\left(t\right) = \left [y_0\left( t\right),y_1\left( t\right),\cdots,y_{K-1}\left( t\right)  \right ]^{T},
    \end{aligned}
    \end{equation}
    \begin{equation}
    \begin{aligned}
    x_n\left( t\right)= \frac{\gamma}{d_r}\frac{\alpha_n}{d_{s,n}}s\left (t - \tau_{s,n} -\tau_r \right ),
    \end{aligned}
    \end{equation}  
 
    \begin{equation}
    \begin{aligned}
    \mathbf{x}\left(t\right) = \left [x_0\left( t\right),x_1\left( t\right),\cdots,x_{N-1}\left( t\right)  \right ]^{T},
    \end{aligned}
    \end{equation}
    \begin{equation}
    \begin{aligned}
    \mathbf{v}\left(t\right) = \left [v_0\left( t\right),v_1\left( t\right),\cdots,v_{K-1}\left( t\right)  \right ]^{T},
    \end{aligned}
    \end{equation}
 
    \begin{equation}
    \begin{aligned}
    \label{eqn:G}
    \mathbf{G} = \begin{bmatrix}
    g_{0,0} &g_{0,1}&\cdots & g_{0,M-1}\\
    g_{1,0}& g_{1,1}& \cdots & g_{1,M-1}\\
    \vdots & \vdots & \ddots & \vdots\\
    g_{K-1,0} &g_{K-1,1} &\cdots & g_{K-1,M-1}
    \end{bmatrix},
    \end{aligned}
    \end{equation}
where $g_{k,m}=A_{k,m}e^{j\phi_{k,m}}$, define $\mathbf{a}\!\left ( \theta \right ) = \left [ 1, e^{j2\pi d \text{sin}{(\theta +\varphi) }}, \cdots, e^{j2\pi \left( M-1\right) d \text{sin}{(\theta +\varphi) }} \right ]^{\text{T}} $ as the steering vector and the array manifold matrix $\mathbf{A}\!\left ( \bm{\theta} \right ) = \left[ \mathbf{a}\!\left ( \theta_0 \right ), \mathbf{a}\!\left ( \theta_1 \right ),\cdots,\mathbf{a}\!\left ( \theta_{N-1} \right )\right]$.

Then, by ignoring $t$ to simplify the notation, the received signal data vector at the receive antenna is
    \begin{equation}
    \begin{aligned}
    \label{eqn:model}
    \mathbf{y} = \mathbf{G} \mathbf{A}\!\left ( \bm{\theta} \right ) \mathbf{x} + \mathbf{v},
    \end{aligned}
    \end{equation}
where $\mathbf{y}\in \mathbb{C}^{K \times 1}$, measurement matrix $\mathbf{G}\in \mathbb{C}^{K \times M}$, $\mathbf{A}(\bm{\theta})\in \mathbb{C}^{M \times N}$, $\mathbf{x}\in \mathbb{C}^{N \times 1}$, and $\mathbf{v}\in \mathbb{C}^{K \times 1}$.

The problem we are interested in solving in this work corresponds to the DOA estimation problem of RIS received signals according to the system model described by equation \eqref{eqn:model}. In other words, the problem is how to obtain an estimate of $\bm{\theta}\;(\bm{\theta} = \left[\theta_0,\theta_1,\cdots,\theta_{N-1}\right ]^{T})$ from $\mathbf{y}$, where $\mathbf{G}$ is known.

\section{Proposed DOA Estimation Algorithm}

In order to get DOA estimate $\bm{\theta}$ from $\mathbf{y}$ and to account for the interference of impulsive-noises, a robust approach is created on the basis of the Lawson norm, which uses the Lawson norm as the error criterion and  a regularization constraint. Fig.\,\ref{fig:norm_vs} shows several typical curves of objective functions: Lawson norm's criterion, the $l_1$ norm, Huber's criterion\cite{Huber_1,Huber_2} and the $l_2$ norm. It can be seen that the $l_2$ norm is very sensitive to outliers and therefore unsuitable for the situation considered in this paper. In contrast, the Lawson norm has a relatively flat curve in the large error range, making it effectively suppress outlier effects. Compared to the Huber's and $l_1$ norm's criterion, it is more sensitive to small errors and can achieve fast convergence.
    \begin{figure}[t]
    \centering
    \includegraphics[width=3.5in]{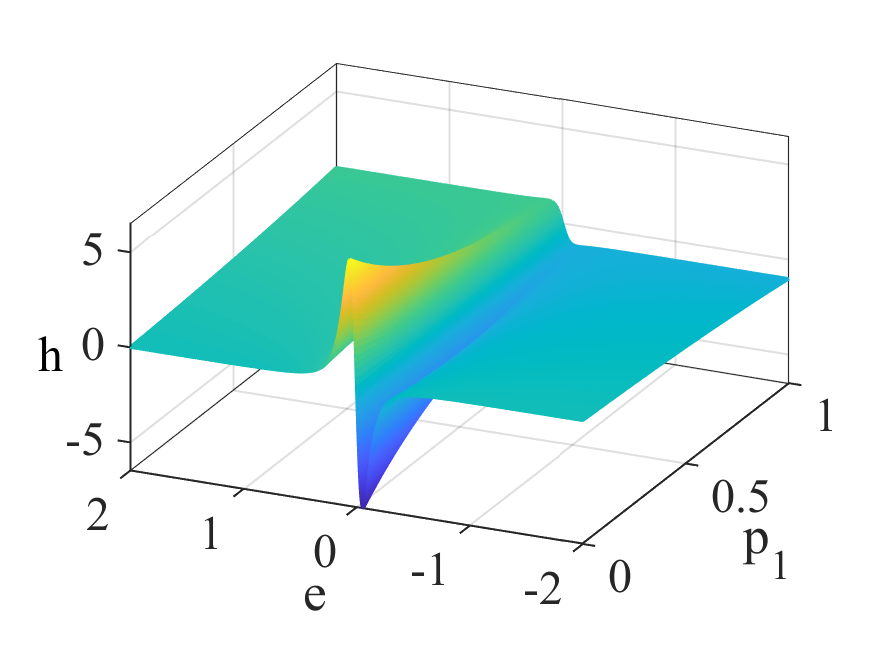}%
    \caption{The 3D image of the derivative of the Lawson norm ($\lambda = 0.1 $).}
    \label{fig:3D_LN}
    \end{figure}

The estimation error constraint and the signal variation sparse reconstruction based on the Lawson norm are expressed respectively as
    \begin{equation}
    \begin{aligned}
    \left \| \mathbf{e} \right \|_{La_1} = \sum_{k=0}^{K-1} \frac{\left |e\left(k\right)\right |^2}{\left(\left |e\left(k\right)\right |^2 + \lambda_1^2\right)^{\frac{2-p_1}{2}}},
    \end{aligned}
    \end{equation}
    \begin{equation}
    \begin{aligned}
    \left \| \mathbf{z} \right \|_{La_2} = \sum_{m=0}^{M-1} \frac{\left |z\left(m\right)\right |^2}{\left(\left |z\left(m\right)\right |^2 + \lambda_2^2\right)^{\frac{2-p_2}{2}}},
    \end{aligned}
    \end{equation}
where $\lambda_1, \lambda_2$ and $p_1, p_2$ are the parameters of the Lawson norm, $\mathbf{z}=\mathbf{A} \left( \mathbf{ \theta} \right ) \mathbf{x}$ denotes the reconstructed signal and $\mathbf{e}=\mathbf{y} - \mathbf{G}\mathbf{z}$ denotes the estimation error. Fig.\,\ref{fig:3D_LN} further describes the properties of the Lawson norm's objective function, for $p<1$, which can handle outliers caused by impulsive noises.

The proposed optimization problem based on the dual Lawson norm is expressed as follows:
    \begin{equation}
    \begin{aligned}
    \label{eqn:question}
    \min_{\mathbf{e},\mathbf{z}} &\left \| \mathbf{e} \right \|_{La_1} + \rho \left \| \mathbf{z} \right \|_{La_2}
    \\&\text{s.t.} \enspace \mathbf{e} = \mathbf{y} - \mathbf{G}\mathbf{z}
    \end{aligned},
    \end{equation}
where $\rho > 0$ is a regularization-parameter that trade-offs the estimation error and sparsity of reconstructed signal. The value of $\rho$ largely determines the algorithm performance. Usually, optimal value of $\rho$ is chosen based on the noise $\mathbf{v}$, the structure of the optimization problem, and the true signal $\mathbf{y}$\cite{L.2021}.

The above optimization problem is solved to obtain the reconstructed signal $\mathbf{z}$, which contains the information of DOA $\bm{\theta}$. Since the targets in spatial domain is sparse, the MUSIC algorithm is used to get spatial spectrum of $\mathbf{z}$, with the positions of the spectral peaks being the desired DOAs $\bm{\theta}$. This paper uses an ADMM method combined with Bregman distance\cite{Bregman2010} to solve the optimization problem~\eqref{eqn:question}, and the solution process is explained in detail below.

Firstly, let us rewrite~\eqref{eqn:question} as an augmented Lagrangian function~\cite{Lshu2014} as
    \begin{equation}
    \begin{aligned}
    \label{eqn:Lag}        
    \mathcal{L}_\epsilon(\mathbf{z},\mathbf{e}, \bm{\xi}) = \left \| \mathbf{e} \right \|_{La_1} + \rho \left \| \mathbf{z} \right \|_{La_2} + \left \langle \mathbf{e}-\mathbf{y} + \mathbf{G} \mathbf{z},\bm{\xi} \right \rangle + \frac{\epsilon}{2} \left \| \mathbf{e}-\mathbf{y} + \mathbf{G} \mathbf{z}\right \|_2^2,
    \end{aligned}
    \end{equation}
where $\bm{\xi}\in\mathbb C^{K \times 1}$ is the introduced dual variable, commonly known as the augmented Lagrangian multiplier, and $\epsilon$ is a parameter that controls the penalty weight. According to equation~\eqref{eqn:Lag}, use the principle of ADMM to obtain iterative expressions for each parameter, as follows:
    \begin{equation}
    \begin{aligned}
    \label{eqn:iter_e_1}
    \mathbf{e}^{i+1}= \text{arg} &\min_{\mathbf{e}}  \Big \{ \left \| \mathbf{e}^i \right \|_{La_1}+\left \langle \mathbf{e}^i-\mathbf{y} + \mathbf{G} \mathbf{z}^i,\bm{\xi}^{i} \right \rangle
    \\&+ \frac{\epsilon}{2} \left \| \mathbf{e}^i-\mathbf{y} + \mathbf{G} \mathbf{z}^i\right \|_2^2 + \bigtriangleup_{f_1}(\mathbf{e},\mathbf{e}^{i})\Big \},
    \end{aligned}
    \end{equation}
    \begin{equation}
    \begin{aligned}
    \label{eqn:iter_z_1}           
    \mathbf{z}^{i+1}= \text{arg} &\min_{\mathbf{z}} \Big \{ \rho \left \| \mathbf{z}^i \right \|_{La_2}+\left \langle \mathbf{e}^{i+1}-\mathbf{y} + \mathbf{G} \mathbf{z}^i,\bm{\xi}^{i} \right \rangle
    \\&+ \frac{\epsilon}{2} \left \| \mathbf{e}^{i+1}-\mathbf{y} + \mathbf{G} \mathbf{z}^i\right \|_2^2 + \bigtriangleup_{f_2}(\mathbf{z},\mathbf{z}^{i})\Big \},
    \end{aligned}
    \end{equation}    
    \begin{equation}
    \begin{aligned}
    \label{eqn:iter_xi}                  
    \bm{\xi}^{i+1} = \bm{\xi}^i + \epsilon \left( \mathbf{e}^{i+1}-\mathbf{y}+\mathbf{G} \mathbf{z}^{i+1}\right ),
    \end{aligned}
    \end{equation}
where $\bigtriangleup_{f_i} (i = 1,2)$ is the given Bregman distance\cite{Bregman2010,Wulei2010}, for the convex differential function $f_i$,  its Bregman distance is defined as follows:
    \begin{equation}
    \begin{aligned}
    \label{eqn:Bregman}                  
    \bigtriangleup_{f_i}(x,y) = f_i(x) - f_i(y) - \left \langle \bigtriangledown{f_i}(y),x-y \right \rangle,
    \end{aligned}
    \end{equation}
an appropriate Bregman metric $\bigtriangleup_{f_i}$ can effectively improve the efficiency of the algorithm. In general, let $f_i(x)=\|x\|_2^2$, the Bregman distance~\eqref{eqn:Bregman} is simplified to the Euclidean distance $\|x-y\|_2^2$\cite{Bregman2018}, and in this paper, we also define it in the same way.

Definition $f_i(x)=\frac{\zeta}{2}\|x\|_2^2\,(i = 1,2)$, according to equations~\eqref{eqn:iter_e_1} and~\eqref{eqn:iter_z_1}, we solve problem~\eqref{eqn:question} with the following expressions:
    \begin{equation}
    \begin{aligned}
    \label{eqn:pro_solve}
    \left\{\begin{matrix}
    \mathbf{e}^{i+1}\!= \text{arg} \mathop {\min }\limits_{\mathbf{e}}  \Big\{\left \| \mathbf{e}^i \right \|_{La_1}\!\!+ \frac{\epsilon}{2} \left \| \mathbf{e}^i-\mathbf{u}^i \right \|_2^2  + \frac{\zeta}{2}\|\mathbf{e} - \mathbf{e}^i\|_2^2 \Big\}\\
    \mathbf{z}^{i+1}\! = \text{arg}\mathop {\min }\limits_{\mathbf{z}} \Big \{ \rho\| \mathbf{z}^i \|_{La_2}\! + \frac{\epsilon}{2} \| \mathbf{G}\mathbf{z}^i - \mathbf{t}^{i} \|_2^2 + \frac{\zeta}{2}\|\mathbf{z} - \mathbf{z}^i\|_2^2 \Big \}
    \end{matrix}\right.   
    \end{aligned},
    \end{equation}
where $\mathbf{u}^i = \mathbf{y} - \mathbf{G} \mathbf{z}^i-\frac{\bm{\xi}^{i}}{\epsilon}$, $\mathbf{t}^{i} = \mathbf{y} - \mathbf{e}^{i+1} - \frac{\bm{\xi}^i}{\epsilon}$, and $\zeta$ is a parameter that can be selected based on actual problem.

Obviously, in~\eqref{eqn:pro_solve}, both the $\mathbf{e}$-sub-problem and the $\mathbf{z}$-sub-problem are traditional norm minimization problems, and their solutions can be easily obtained. Fig.\,\ref{fig:3D_LN} shows that when $0\le p_1 <1$, $\left \| \mathbf{e} \right \|_{La_1}$ can handle well the interference of large outliers caused by impulsive noises, but it is a non-convex-problem and is difficult to ensure the convergence to the global solution. In this context, an existing research work \cite{Bregman2018} has demonstrated the global convergence of the Bregman ADMM method for solving non-convex optimization problems. The asymptotic convergence iteration formulas for each parameter are as follows:
    \begin{equation}
    \begin{aligned}
    \label{eqn:last_solve}
    \left\{\begin{matrix}
    &\mathbf{e}^{i+1} = (\mathbf{H}_{e} + (\epsilon+\zeta)\mathbf{E})^{-1}(\epsilon \mathbf{u} + \zeta \mathbf{e}^i)\\
    &\mathbf{z}^{i+1} = (\rho\mathbf{H}_{z}+ (\epsilon+\zeta) \mathbf{G}^{\text{H}}\mathbf{G})^{-1} (\epsilon \mathbf{G}^{\text{H}}\mathbf{t}^{i} + \zeta \mathbf{z}^i)
    \end{matrix}\right.   
    \end{aligned},
    \end{equation}
where $\mathbf{E}$ is a $K$-order identity matrix, $\bigtriangledown _{ \mathbf{e}^{\ast}} \|\mathbf{e}\|_{La_1} = \frac{1}{2}\mathbf{H}_{e}\mathbf{e} $ and $\bigtriangledown_{ \mathbf{z}^{\ast}} \|\mathbf{z}\|_{La_2}=\frac{1}{2}\mathbf{H}_{z}\mathbf{z}$. $\mathbf{H}_e$ and $\mathbf{H}_z$ are both diagonal matrices\cite{self_iceict}, they have the following forms:
    \begin{equation}
    \begin{aligned}
    \label{eqn:H}
    &\mathbf{H}_e = \text{diag} \left [ h_{e0}, h_{e1}, \cdots , h_{e(K-1)} \right ]\\
    &\mathbf{H}_z = \text{diag} \left [ h_{z0}, h_{z1}, \cdots , h_{z(M-1)} \right ]\\
    h_{ek} = &\frac{p_1 \left |e(k)\right |^2 + 2 \lambda_1^2}{\left( \left |e(k)\right |^2 +  \lambda_1^2 \right )^{\frac{4-p_1}{2}}}, k = 0,1,\cdots,K-1\\
    h_{zm} = &\frac{p_1 \left |z(m)\right |^2 + 2 \lambda_1^2}{\left( \left |z(m)\right |^2 +  \lambda_1^2 \right )^{\frac{4-p_1}{2}}}, m = 0,1,\cdots,M-1
    \end{aligned}.
    \end{equation}

So far, we have obtained the iterative equations for all unknown parameters in~\eqref{eqn:question}, which can be used to reconfigure $\mathbf{z}$ from $\mathbf{y}$.

The paper is to perform DOA estimation for wireless communication systems, and the DOA $\bm{\theta}$ can be obtained from the reconstructed signal $\mathbf{z}$ using the MUSIC algorithm. The standard MUSIC algorithm \cite{music} cannot be used for getting DOA-estimation with single snapshot, and a MUSIC algorithm based on a Hankel matrix is used in this paper~\cite{wliao2015}.

Let us now describe the MUSIC-type algorithm that employs a Hankel matrix and how it is used in the proposed LN-MUSIC algorithm. First, let us construct the Hankel matrix of the reconstructed signal $\mathbf{z}$ in the form of
    \begin{equation}
    \label{eqn:hankel}
    \text{Hankel}\left(\mathbf{z} \right)=\begin{bmatrix}
    z_0 & z_1 & \cdots & z_{M-L}\\
    z_1 & z_2 & \cdots & z_{M-L+1}\\
    \vdots & \vdots & \ddots & \vdots\\
    z_{L-1} & z_{L} & \cdots & z_{M-1}
    \end{bmatrix},
    \end{equation}
where $L$ must satisfy $N\le L < M-N+1$, and it is a positive integer. Suppose the singular values of $\text{Hankel}(\mathbf{z})$ are $\delta_0, \delta_1, \cdots, \delta_{N-1}$ $(\delta_0 > \delta_1 > \cdots > \delta_{N-1} > 0)$, and then perform singular value decomposition (SVD) on matrix~\eqref{eqn:hankel} is
    \begin{equation}
    \begin{aligned}
    \text{Hankel} \left( \mathbf{z} \right ) = \left[\mathbf{U}_1 \enspace \mathbf{U}_2 \right ] \begin{bmatrix}
    \Delta & \mathbf{0}\\
    \mathbf{0} & \mathbf{0}       
    \end{bmatrix} \mathbf{V}^{\rm{H}}   
    \\ \Delta =\text{diag} \left( \delta_0, \delta_1, \cdots, \delta_{N-1} \right )      
    \end{aligned},
    \end{equation}
where $\mathbf{U}_1 \in \mathbb C^{L \times N}$ and $\mathbf{U}_2 \in \mathbb C^{L \times (L-N)}$ are unitary matrices formed by the left singular vectors that is corresponding to singular values, and their column spaces respectively represent the subspace of signals and the subspace of noise. $\mathbf{V}\in \mathbb C^{(M-L+1) \times (M-L+1)}$ is also a matrix that satisfies the unitary property and is composed of the right singular vectors associated with singular values.

Steering vector is defined as
    \begin{equation}
    \begin{aligned} 
    \mathbf{a}\!\left ( \theta_n \right ) = \left [ 1, e^{j2\pi d \text{sin}{(\theta_n +\varphi) }}, \cdots, e^{j2\pi \left( L-1\right) d \text{sin}{(\theta_n +\varphi) }} \right ]^{\rm{T}},
    \end{aligned}
    \end{equation}
where it is known that the noise and the target signal are mutually independent. Thus, we can state that each steering vector $\mathbf{a}\!\left ( \theta_n \right )$ is orthogonal to the noise subspace $\mathbf{U}_2$, that is
    \begin{equation}
    \begin{aligned} 
    \mathbf{a}\!\left ( \theta_n \right )^{H} \mathbf{U}_2 = 0, n=0,1,\cdots,N-1.
    \end{aligned}
    \end{equation}
        
We then get the spatial spectrum of the reconstructed target signal $\mathbf{z}$ as follows:  
    \begin{equation}
    \begin{aligned} 
    {\varUpsilon\left(\theta\right)}=\frac{\left \|  \mathbf{a}\!\left ( \theta \right )\right\|_2^2}{\left\|\mathbf{a}^{H}\!\left ( \theta \right )\mathbf{U}_2\right\|_2^2},
    \end{aligned}
    \end{equation}
where each target signal DOA $\theta_n$ corresponds to the spectral peak position.

In addition, to mitigate the performance degradation caused by path loss uisng LN-MUSIC algorithm, we propose a strategy called ROSM to optimize the control matrix of the RIS. According to~\eqref{eqn:G}, generate $\bm{\phi}^t (t = 1,2,\cdots,T)$ randomly, and where
    \begin{equation}
    \begin{aligned}
    \bm{\phi}^t = \begin{bmatrix}
    \phi_{0,0}^t &\phi_{0,1}^t&\cdots & \phi_{0,M-1}^t\\
    \phi_{1,0}^t& \phi_{1,1}^t& \cdots & \phi_{1,M-1}^t\\
    \vdots & \vdots & \ddots & \vdots\\
    \phi_{K-1,0}^t &\phi_{K-1,1}^t &\cdots & \phi_{K-1,M-1}^t
    \end{bmatrix}
    \end{aligned},
    \end{equation}
for each $\bm{\phi}^t$, there is a corresponding control matrix $\mathbf{G}^t$. The aim of ROSM is to find the optimal $\mathbf{G}$ to maximize receiving signal power (SRP) 
    \begin{equation}
    \begin{aligned}
    &\rm{SRP} = \frac{\| \mathbf{y} \|_2^2}{\mathit{K}}\\
    \rm{the \enspace optimal\enspace} &\mathbf{G} = \rm{arg} \max_{\mathbf{G}^{\mathit{t}}}\,\rm{SRP}, 1\le \mathit{t}\le \mathit{T}
    \end{aligned}.
    \end{equation}

Finally, Algorithm 1 outlines the main procedure for the simulation implementation of the proposed LN-MUSIC.
    \begin{table}[htbp]   
    \centering
    \begin{tabular}{lcl}    
    \toprule    
    \textbf{Algorithm 1} {The Proposed LN-MUSIC Algorithm}\\   
    \midrule   
    1:\textbf{Input:} antenna receive data vector $\mathbf{y}$, number of RIS reflection\\ units $M$, number of targets $N$, maximum iterations $i_{\text{num}}$. \\ 
    2:\textbf{Initialization:} Set the initial values of some major parameters \\$\mathbf{u}=\mathbf{y}-\mathbf{G}\mathbf{z}+\frac{\bm{\xi}}{\epsilon}, \mathbf{t} = \mathbf{y}-\mathbf{e}+\frac{\bm{\xi}}{\epsilon}, q_1 = 0.4, q_2=1, \lambda_1=0.3,$\\$ \lambda_2=0.001, \mathbf{z} = \mathbf{0},\bm{\xi}=\mathbf{0}, i=0$, $i_{num} = 40$. \\ 
    3: \textbf{Repeat}\\
    \quad 4: Execute ROSM to optimization matrix $\mathbf{G}$:\\
    \quad 5: Set $\mathbf{H}_e$ and
    $\mathbf{H}_z$ according to~\eqref{eqn:H}\\
    \quad 6: Update $\xi$: $\bm{\xi}^{i+1} = \bm{\xi}^i + \epsilon \left( \mathbf{e}^{i+1}-\mathbf{y}+\mathbf{G} \mathbf{z}^{i+1}\right )$\\
    \quad 7: Update $\mathbf{e}$: $\mathbf{e}^{i+1} = (\mathbf{H}_{e} + (\epsilon+\zeta)\mathbf{E})^{-1}(\epsilon \mathbf{u} + \zeta \mathbf{e}^i)$\\
    \quad 8: Update $\mathbf{z}$: $\mathbf{z}^{i+1} = (\rho\mathbf{H}_{z}+ (\epsilon+\zeta) \mathbf{G}^{\text{H}}\mathbf{G})^{-1} \epsilon \mathbf{G}^{\text{H}}\mathbf{t}^{i}$\\
    \quad 9: Update $\lambda_1$:\\
    \qquad 10: if $\lambda_1 > 0.01$\\
    \qquad \quad $\lambda_1 = \eta \lambda_1$, where $\eta \in (0,1)$\\
    \qquad 11: else\\
    \qquad \quad set $\lambda_1 = 0.01$\\
    \qquad 12: end\\
    13: \textbf{Until} $i>i_\text{num}$\\
    14: \textbf{Output:} $\mathbf{z}$.\\
    15: Using the Hankel-based MUSIC algorithm, the spatial-spectrum\\ of $\mathbf{z}$ is $\mathbf{J}\left(\theta\right)=\frac{\left \|  \mathbf{a}\!\left ( \theta \right )\right\|_2^2}{\left\|\mathbf{a}^{H}\!\left ( \theta \right )\mathbf{U}_2\right\|_2^2}$.\\
    16: Perform spectral peak detection, get the spatial spectral peak\\ position of $\mathbf{z}$.\\ 
    \textbf{Output:} DOA $\bm{\theta}$.\\
    \bottomrule   
    \end{tabular}  
    \end{table}

\section{The CRLB for DOA Estimation with Impulsive Noises}

In this section, we detail the modeling of impulsive noises and derive the CRLB for DOA estimation with impulsive noises.

\subsection{System Noise}

This paper discusses the performance of the algorithm with impulsive noise. Impulsive noise~\cite{LC2022} has the characteristics of sudden impulsive amplitude, short duration, etc. The mixed Gaussian noise model\cite{MGN2000} is generally used to describe this noise, and its expression is
    \begin{equation}
    \begin{aligned} 
    v \left( t \right) = p \left ( t \right) \omega_1(t) + \omega_2 \left( t \right),
    \end{aligned}
    \end{equation}
where both $\omega_1(t)$ and $\omega_2(t)$ are additive Gaussian white noise (AGWN), and their variances are represented by $\sigma_{1}^2$ and $\sigma_{2}^2$ $(\sigma_{1}^2 = 100\sigma_{2}^2)$, respectively. $p(t)$ is Bernoulli process, with only $0$ and $1$ values, which can simulate the sudden jump characteristics of impulsive noises.

It is worth noting that the occurrence probability of $\omega_1(t)$ is very small $(\approx 0.1)$, so the impulsive noise model can also be considered as the insertion of several large outliers into the Gaussian-noise background. The noise variance is $(\kappa \sigma_1^2 + \sigma_2^2)$, where $\kappa$ is a probability of the occurrence of the Bernoulli-event, and $f(v(t)) \sim \mathcal{N}(0,\kappa \sigma_1^2 + \sigma_2^2) $. In the following simulations, we define the SNR of the system
    \begin{equation}
    \text{SNR} = \frac{\|\mathbf{z}\|_2^2}{\sigma_{2}^2}.
    \end{equation}

It is worth noting that since RIS does not amplify noise\cite{RISnoise2022}, it can instead enhance the signal of interest and improve the SNR via adjusting the control matrix of RIS. Therefore, in the simulation examples, the SNR is set based on the reconstructed signal $\mathbf{z}$ received by RIS.

\subsection{The CRLB for DOA Estimation with Impulsive Noises}
    
The CRLB\cite{CRLB1,CRLB2} can be used to benchmark the performance of DOA-estimation algorithms. Here, the theoretical CRLB for DOA-estimation with impulsive noises is presented, and its specific procedures are given as follows.

Firstly,~\eqref{eqn:model} contains two unknown parameters $\mathbf{\theta}$ and $\mathbf{x}$, and we collect them into a vector
    \begin{equation}
    \begin{aligned}
    \mathbf{\Gamma} \buildrel \Delta \over = \left [\mathbf{\theta}^{\rm{T}},
    \mathbf{x}^{\rm{T}} \right]^{\rm{T}}     
    \end{aligned},
    \end{equation}
    
Then, the log likelihood function of the received signal $\mathbf{y}$ with respect to $\mathbf{\Gamma}$ is
    \begin{equation}
    \begin{aligned}
    \ln{f(\mathbf{y}\mid\mathbf{\Gamma})} =\, -(\mathbf{y} - \mathbf{G}\mathbf{A}(\mathbf{\theta}) \mathbf{x})^{\rm{H}} \mathbf{Q}^{-1} (\mathbf{y} - \mathbf{G}\mathbf{A}(\mathbf{\theta}) \mathbf{x})
    - \ln (\pi^K det(\mathbf{Q}))  
    \end{aligned},
    \end{equation}
where ${\rm det}(\cdot)$ is determinant operator of a matrix, while $\mathbf{Q}$ is the covariance matrix of the noise $\mathbf{v}$.  

The Fisher-information matrix is represented as follows
    \begin{equation}
    \begin{aligned}
    \mathbf{F} = \begin{bmatrix}
     E \left[ \frac{\partial \ln^{\rm{H}}\!\!{f(\mathbf{y}\mid\mathbf{\Gamma})}}{\partial \mathbf{\theta}} \frac{\partial \ln{f(\mathbf{y}\mid\mathbf{\Gamma})}}{\partial \mathbf{\theta}} \right ] & E \left[ \frac{\partial \ln^{\rm{H}}\!\!{f(\mathbf{y}\mid\mathbf{\Gamma})}}{\partial \mathbf{\theta}} \frac{\partial \ln{f(\mathbf{y}\mid\mathbf{\Gamma})}}{\partial \mathbf{x}} \right ]\\
     E \left[ \frac{\partial \ln^{\rm{H}}\!\!{f(\mathbf{y}\mid\mathbf{\Gamma})}}{\partial \mathbf{x}} \frac{\partial \ln{f(\mathbf{y}\mid\mathbf{\Gamma})}}{\partial \mathbf{\theta}} \right ]& E \left[ \frac{\partial \ln^{\rm{H}}\!\!{f(\mathbf{y}\mid\mathbf{\Gamma})}}{\partial \mathbf{x}} \frac{\partial \ln{f(\mathbf{y}\mid\mathbf{\Gamma})}}{\partial \mathbf{x}} \right ]
    \end{bmatrix}  
    \end{aligned},
    \end{equation}
where $E\left[\cdot\right]$ is the mathematical expectation operator, we have
    \begin{equation}
    \begin{aligned}
    \frac{\partial \ln{f(\mathbf{y}\mid\mathbf{\Gamma})}}{\partial \bm{\theta}} &= -\frac{\partial (\mathbf{y} - \mathbf{G}\mathbf{A}(\mathbf{\theta}) \mathbf{x})^{\rm{H}} \mathbf{Q}^{-1} (\mathbf{y} - \mathbf{G}\mathbf{A}(\mathbf{\theta}) \mathbf{x})}{\partial \bm{\theta}}\\
    & = 2 Re\left[ (\mathbf{y} - \mathbf{G}\mathbf{A}(\mathbf{\theta}) \mathbf{x})^{\rm{H}} \mathbf{Q}^{-1} \mathbf{G} \frac{\partial \mathbf{A}(\mathbf{\theta}) \mathbf{x}}{\partial \bm{\theta}}\right]
    \end{aligned},
    \end{equation}
    \begin{equation}
    \begin{aligned}
    \frac{\partial \ln{f(\mathbf{y}\mid\mathbf{\Gamma})}}{\partial \mathbf{x}} &= -\frac{\partial (\mathbf{y} - \mathbf{G}\mathbf{A}(\mathbf{\theta}) \mathbf{x})^{\rm{H}} \mathbf{Q}^{-1} (\mathbf{y} - \mathbf{G}\mathbf{A}(\mathbf{\theta}) \mathbf{x})}{\partial \mathbf{x}}\\
    & = (\mathbf{y} - \mathbf{G}\mathbf{A}(\mathbf{\theta}) \mathbf{x})^{\rm{H}} \mathbf{Q}^{-1} \mathbf{G} \mathbf{A}(\mathbf{\theta})
    \end{aligned},
    \end{equation}
where a similar expression to $\frac{\partial \mathbf{A}(\mathbf{\theta}) \mathbf{x}}{\partial \bm{\theta}}$ was reported in \cite{L2_ANM_CVX}, and we adapt the algebraic structure in $\mathbf{B} = \frac{\partial \mathbf{A}(\mathbf{\theta}) \mathbf{x}}{\partial \bm{\theta}}$ from reference\cite{L2_ANM_CVX}. Therefore, the Fisher matrix is represented as
    \begin{equation}
    \begin{aligned}
    \label{eqn:Fisher_1}
    \mathbf{F} = \frac{1}{\kappa \sigma_1^2 + \sigma_2^2}\begin{bmatrix}2Re\left[\mathbf{B}^{\rm{H}}\mathbf{G}^{\rm{H}}\mathbf{G}\mathbf{B} \right] & \mathbf{B}^{\rm{H}}\mathbf{G}^{\rm{H}}\mathbf{G}\mathbf{A}\\
   \mathbf{A}^{\rm{H}}\mathbf{G}^{\rm{H}}\mathbf{G}\mathbf{B}& \mathbf{A}^{\rm{H}}\mathbf{G}^{\rm{H}}\mathbf{G}\mathbf{A} 
    \end{bmatrix} .   
    \end{aligned}
    \end{equation}

{Let the four sub-blocks of matrix $\mathbf{F}$ be represented by $\mathbf{F}_{11}$, $\mathbf{F}_{12}$, $\mathbf{F}_{21}$ and $\mathbf{F}_{22}$, respectively, hence \eqref{eqn:Fisher_1} can be rewritten as
    \begin{equation}
    \begin{aligned}
    \mathbf{F} = \begin{bmatrix}\mathbf{F}_{11} & \mathbf{F}_{12}\\
   \mathbf{F}_{21}& \mathbf{F}_{22} 
    \end{bmatrix} .
    \end{aligned}
    \end{equation}}

{The CRLB can be derived from the inverse of the Fisher Information Matrix. It is known that the parameter $\bm{\theta}$ is the first element of $\bm{\Gamma}$. According to the definition of the inverse of a block matrix, the CRLB for the desired DOA can be expressed as follows\cite{CRLB_Fisher}:
    \begin{equation}
    \begin{aligned}
    \text{CRLB} = \left[ \mathbf{F}_{11} - \mathbf{F}_{12}\mathbf{F}_{22}^{-1}\mathbf{F}_{21}^{\rm{T}}\right]^{-1}   
    \end{aligned}.
    \end{equation}}

\section{Simulations and Discussions}
    
In this section, we assess the {LN-MUSIC} using simulation experiments that were carried out on a personal computer using MATLAB R2021b software. Firstly, we explored the impact of various parameters on the performance of LN-MUSIC to identify the optimal settings. Subsequently, the {LN-MUSIC} is compared with other existing methods with respect to computational complexity, resolution, estimate accuracy and other aspects. A series of simulation experiments have verified the superiority of the {LN-MUSIC}. The system parameter settings are given in Table\,\ref{tab:parameters}.

    \begin{table}[!htbp]
    \centering
    \caption{Explanation of simulation symbols} 
    \label{tab:parameters} 
    \begin{tabular}{ccc} 
    \toprule 
     Symbols $\!\!\!$& Interpretation $\!\!\!\!\!\!$& Value \\
    \midrule 
     $K$ $\!\!\!$& Time slots $\!\!\!\!\!\!$&  128 \\
     $M$ $\!\!\!$& Number of RIS elements $M$$\!\!\!\!\!\!$& 32 \\
     \makecell[c]{$d$} & \makecell[c]{Wavelength-normalized distance\\ between adjacent RIS elements} $\!\!\!\!\!\!$& 0.5 \\
     $N$ $\!\!\!$& Targets Number $\!\!\!\!\!\!$& 3 \\
     $d_{r}$ $\!\!\!$& Distance separating RIS from the targets$\!\!\!\!\!\!$& $30 \,\text{m}$\\
     $d_{s}$ $\!\!\!$& Distance separating RIS from the antenna$\!\!\!\!\!\!$& $3\,\text{m}$\\
     $\alpha_n, \gamma$ & Path loss constant $\!\!\!\!\!\!$& $1$\\
     $i_{\text{num}}$ $\!\!\!$& Iterations $\!\!\!\!\!\!$& 30 \\
     $\varphi$ $\!\!\!$&Azimuth angle between RIS and antenna $\!\!\!\!\!\!$& $10^{\circ}$\\
     $\theta_{\text{range}}$ $\!\!\!$& Range of spatial angle $\!\!\!\!\!\!$&$\!\!\!\!\!\! \left (-40^{\circ},40^{\circ}\right )$\\
     $\bm{\theta}$ $\!\!\!$& Mean value of DOA & $\!\!\!\!\!\!\left [-20^{\circ},0^{\circ},25^{\circ} \right ]$\\
    \bottomrule 
    \end{tabular}
    \end{table}

In order to validate the behavior of the presented LN-MUSIC and to make a comparison with several existing approaches, this paper introduces the root-mean square-of-error (RMSE), giving by
    \begin{equation}
    \text{RMSE}=\sqrt{\frac{1}{NN_{\text{MC}}} \sum_{i_\text{MC}=0}^{N_{\text{MC}}-1} \left \| \hat{\bm{\theta}}_{i_\text{MC}} - \bm{\theta}_{i_\text{MC}} \right \|_2^2},
    \end{equation}
where $N_{\text{MC}}$ is the number of Monte Carlo experiments, $\hat{\bm{\theta}}$ denotes the DOA achieved through the proposed algorithm, while $\bm{\theta}$ refers to the actual DOA. The simulation results are discussed in detail below.

\begin{figure}[h]
    \centering
    \includegraphics[width=3.5in]{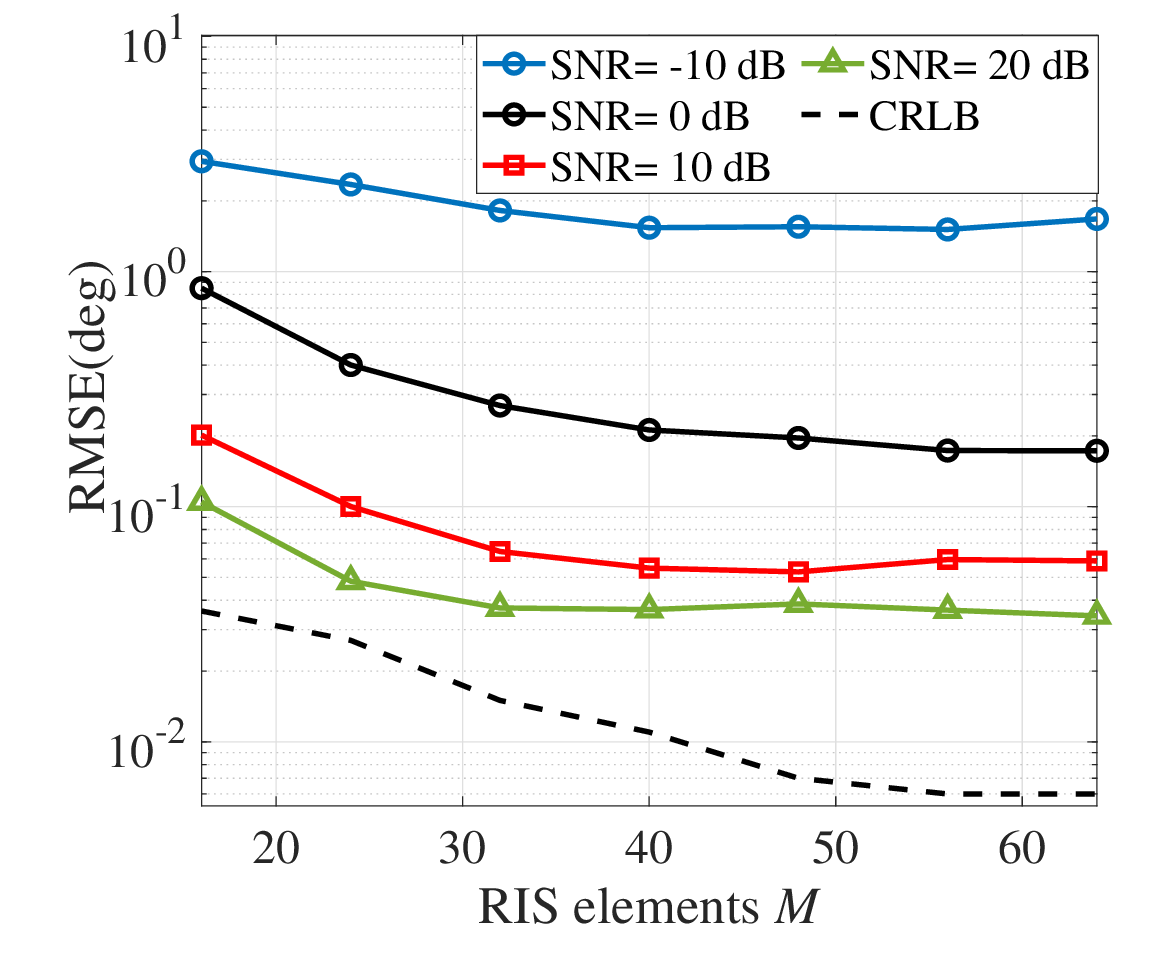}%
    \caption{The RMSE of {LN-MUSIC} with different $M$ $(K=128)$.}
    \label{fig:Param.M}
    \end{figure}

\begin{figure}[h]
    \centering
    \includegraphics[width=3.5in]{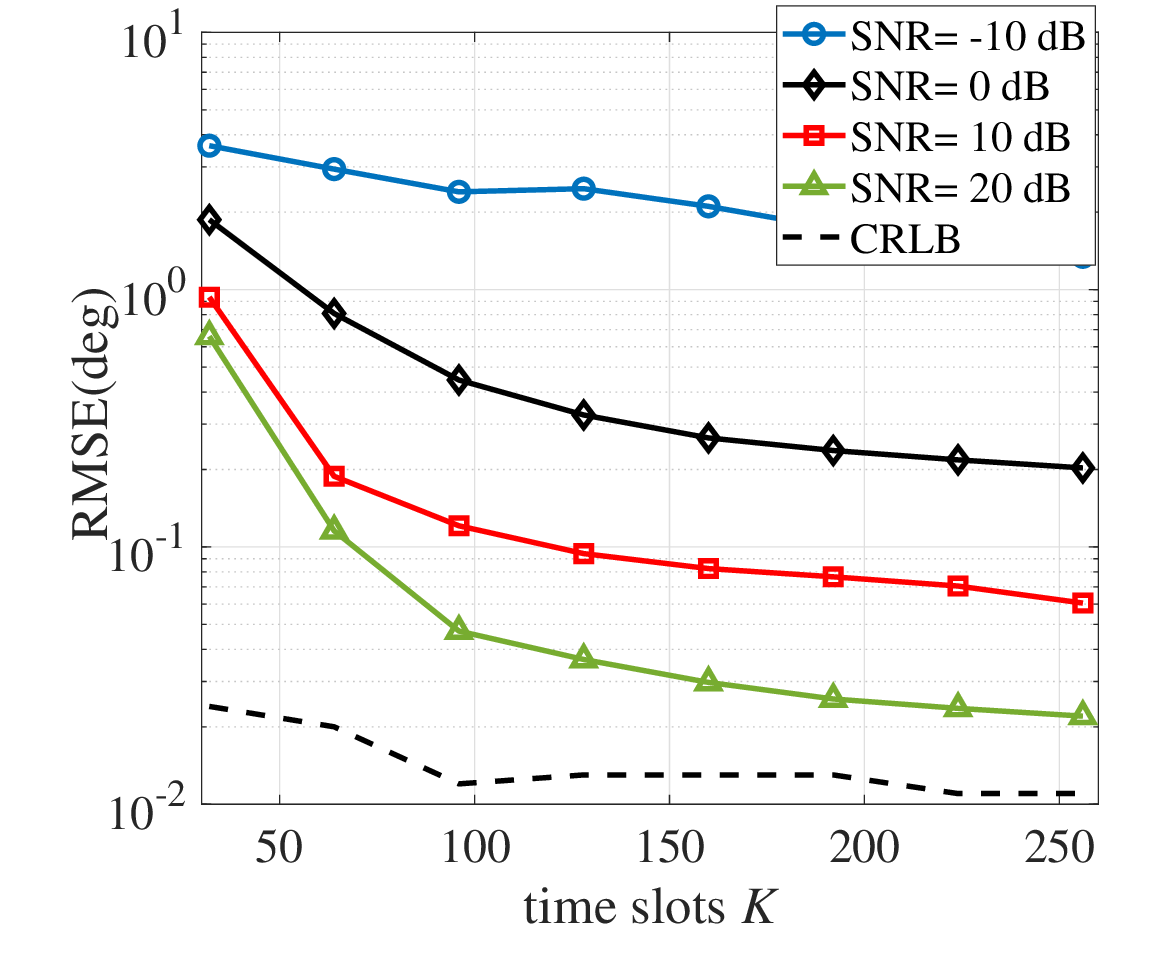}%
    \caption{The RMSE of {LN-MUSIC} with different $K$ $(M=32)$.}
    \label{fig:Param.K}
    \end{figure}

For the DOA estimation model considered in this work, $M$ is a very important factor that influences the results. Many existing studies demonstrated that as $M$ increases, the performance of the DOA-estimation algorithm also improves. Fig.\,\ref{fig:Param.M} validates the effect of $M$ on the {LN-MUSIC}'s accuracy with various SNRs. We find that a larger $M$ corresponds to a higher accuracy, i.e. RMSE. However, when $M>30$, the curve becomes flatter as $M$ increases. The reason is that this paper adopts the ULA structure of RIS, and when $M$ increases to a certain level, the size of the RIS becomes too large, resulting in blind spots and unable to form an effective reflection path. In addition, Fig.\,\ref{fig:Param.K} gives the effect of $K$ on estimation accuracy of the {LN-MUSIC}. Similarly, when $K$ increases to a certain level, the improvement in accuracy of the algorithm is limited. Therefore, in all the following simulations, we set the parameters to $K=128$ and $M=32$.

{We have taken into account the impact of the path loss on the algorithm performance during the system modeling. The RIS can perform beamforming to enhance the power of the reflected signals, hence, the path between the RIS and the receiver has the most pronounced effect on the algorithm. In Fig.\,\ref{fig:Param.ds}, we demonstrate the influence of $d_s$ on the estimation accuracy. According to \eqref{eqn:ds_y}, as $d_s$ increases, the power of the signal $\mathbf{y}$ will continue to decrease, which is detrimental to the operation of the DOA estimator. From Fig.\,\ref{fig:Param.ds}, it can be observed that under different SNR, as $d_s$ increases, the RMSE also increases, which is consistent with the aforementioned inference. However, we can observe that the rise in the curve is not significant, especially when the SNR is greater than 10 dB, the accuracy of the proposed LN-MUSIC estimator can be maintained within a certain range $(< 0.1^{\circ})$, demonstrating the robustness of the LN-MUSIC estimator.
    \begin{figure}[htbp]
    \centering
    \includegraphics[width=3.5in]{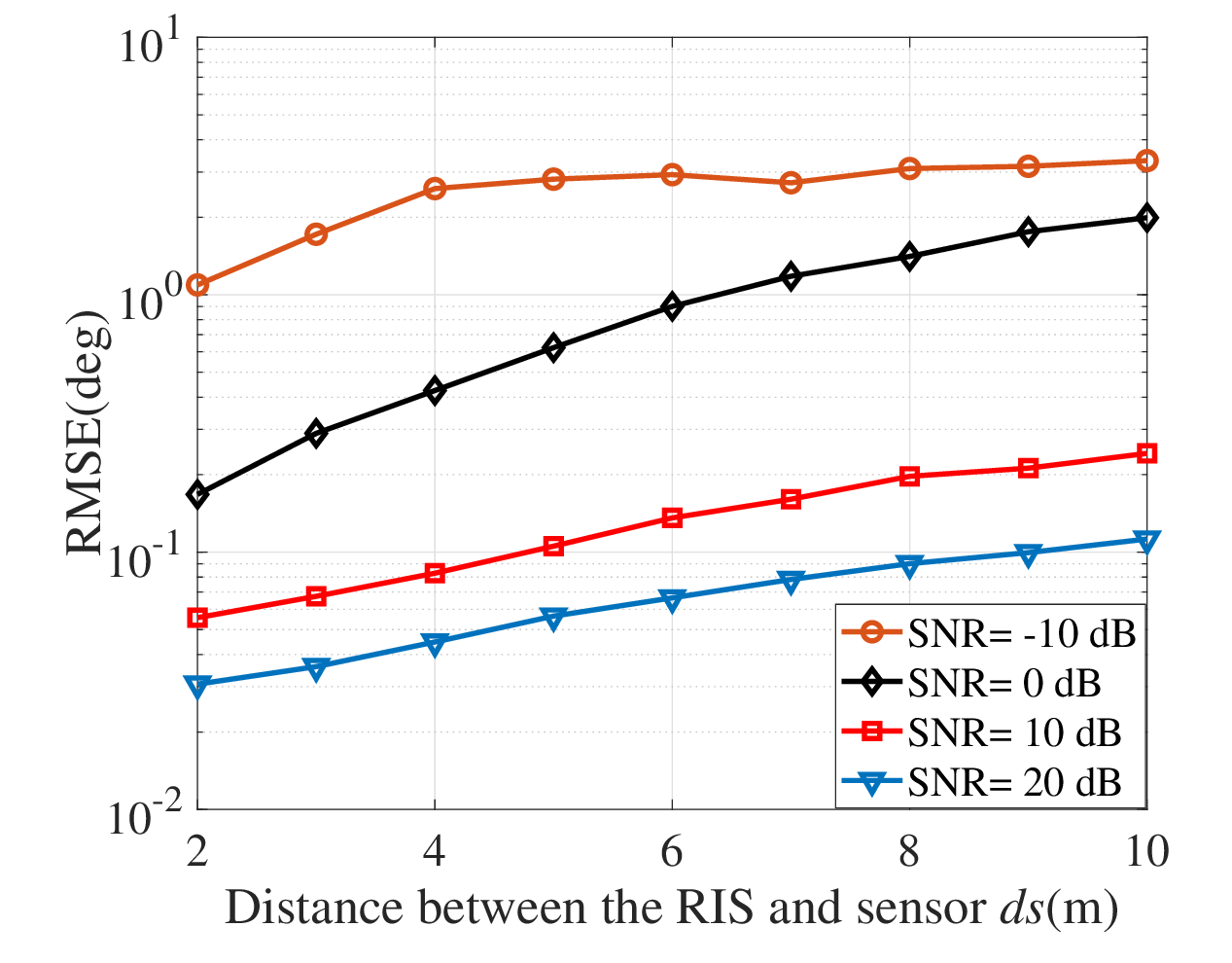}%
    \caption{The RMSE of {LN-MUSIC} with different $ds$ $(dr=30\,\rm{m})$.}
    \label{fig:Param.ds}
    \end{figure}}

\begin{figure}[htbp]
    \centering
    \includegraphics[width=3.5in]{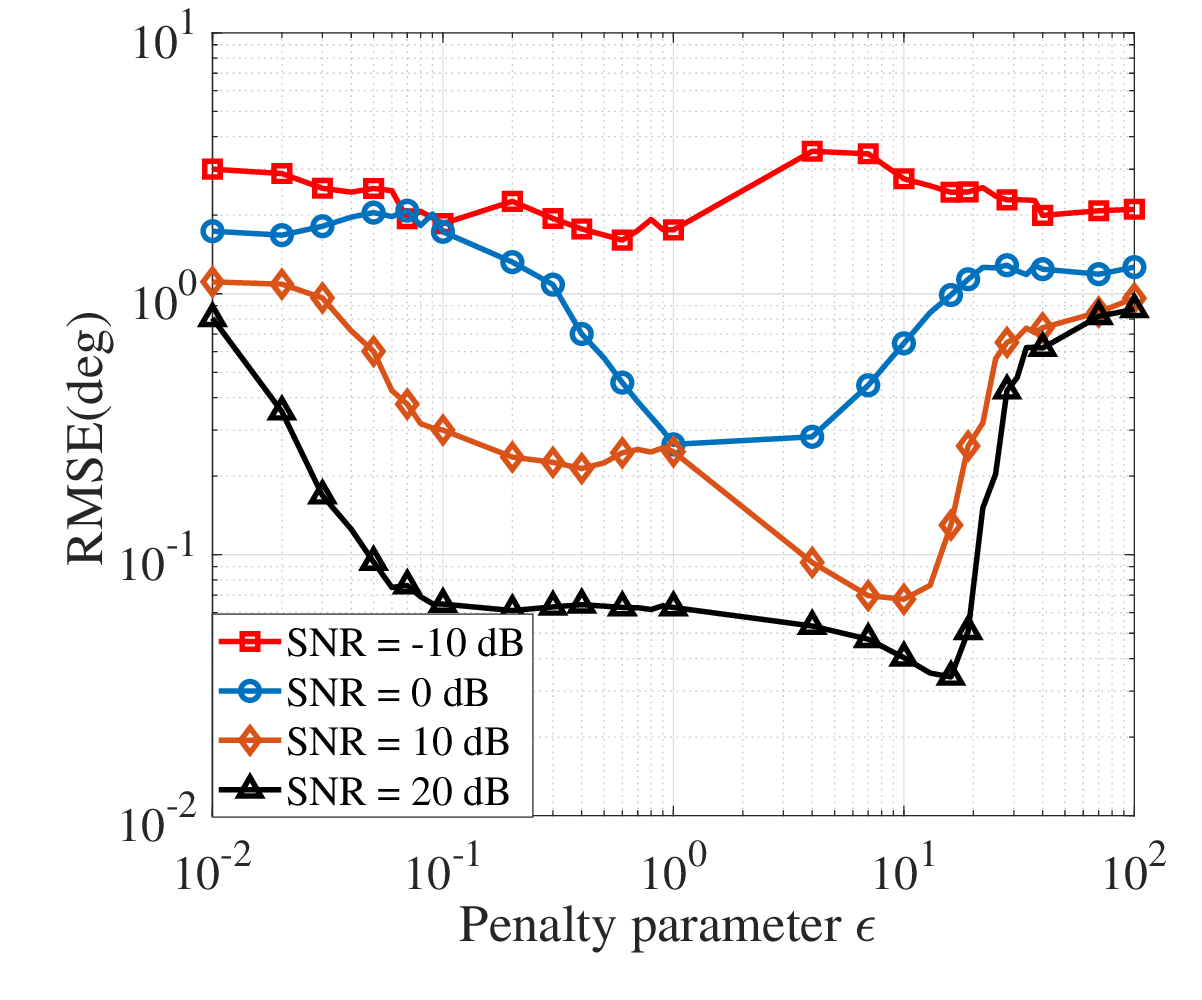}%
    \caption{The RMSE of {LN-MUSIC} with different $\bm{\epsilon}$.}
    \label{fig:eps}
    \end{figure}

\begin{figure}[htbp]
    \centering
    \includegraphics[width=3.5in]{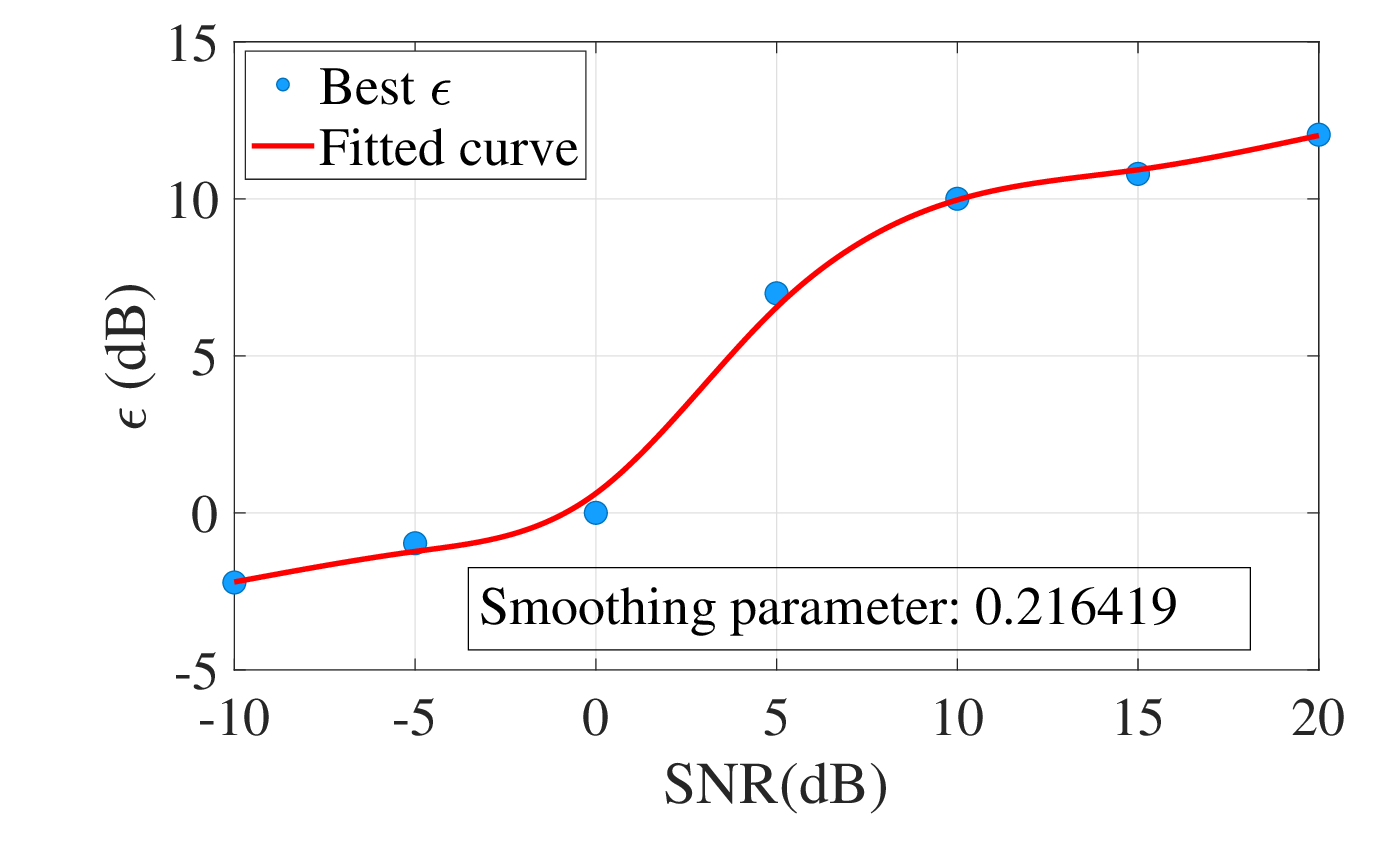}%
    \caption{The optimal value curve of $\bm{\epsilon}$ with different SNRs.}
    \label{fig:eps_cur}
    \end{figure}

Since in this work we employ a Bregman ADMM method to find out a solution of this formulated non-convex-optimization problem, choosing appropriate penalty parameters $\bm{\epsilon}$ is the key to achieving good performance \cite{eps2017}. However, currently there is no general method to guide the finding of the optimal $\bm{\epsilon}$. Therefore, this paper searches for the optimal $\bm{\epsilon}$ corresponding to different SNRs through parameter scanning and nonlinear fitting. The results of parameter scanning are shown in Fig.\,\ref{fig:eps} for SNRs v.s. $\bm{\epsilon}$. Based on the information in Fig.\,\ref{fig:eps}, nonlinear fitting was performed on the optimal $\bm{\epsilon}$ corresponding to different SNRs in Fig.\,\ref{fig:eps_cur}, using the smoothing spline method with the smoothing parameter set to 0.21.

In order to verify the superiority of {LN-MUSIC}, a series of comparisons are made between {LN-MUSIC} and other existing methods (OMP\cite{OMP2012}, ANM\cite{Lowcost2022}, Lp-ADM\cite{Lp_MUSIC2013}, SC-ADM\cite{SC_MUSIC2017}, Lp-SEF\cite{SEF2019} and Hub-sniht\cite{Hub_sinht2015}) in terms of accuracy, reliability and computational complexity. {Among the comparative methods provided, the following methods have been modified accordingly based on the system model \eqref{eqn:model}:}

{1) Lp-ADM: consider the optimization problem 
\begin{equation}
\label{eqn:lp_MUSIC}
    \min_{\mathbf{z} \in \mathbb{C}^{N}} \|\mathbf{y} -\mathbf{G}\mathbf{z}\|_p^p + \rho\|\mathbf{z}\|_1
\end{equation}
to reconstruct the desired signal $\mathbf{z}$, solve it using the proximal gradient algorithm\cite{smooth2017}, and finally obtain the DOA information through the MUSIC algorithm. Herein, $p = 0.7$.}

{2) SC-ADM: consistent with the Lp-MUSIC method, the $l_p$-norm error criterion in the objective function \eqref{eqn:lp_MUSIC} is replaced with the Cauchy score function provided in \cite{SC_MUSIC2017}.}

{3) Lp-SEF: the optimization problem
\begin{equation}
\label{eqn:SEF_fun}
    \min_{\mathbf{z} \in \mathbb{C}^{N}} \|\mathbf{y} -\mathbf{G}\mathbf{z}\|_p^p + \rho h_p(\mathbf{z})
\end{equation}
is formulated based on the Shannon entropy function (SEF) regularizer, and the fast iterative shrinkage-thresholding algorithm (FISTA)\cite{FISTA2009} is employed for solving it. Herein, $h_q(\cdot)$ denotes the SEF, and $p = 0.7$, $q = 1.1$.}

\begin{figure}[htbp]
    \centering
    \includegraphics[width=3.5in]{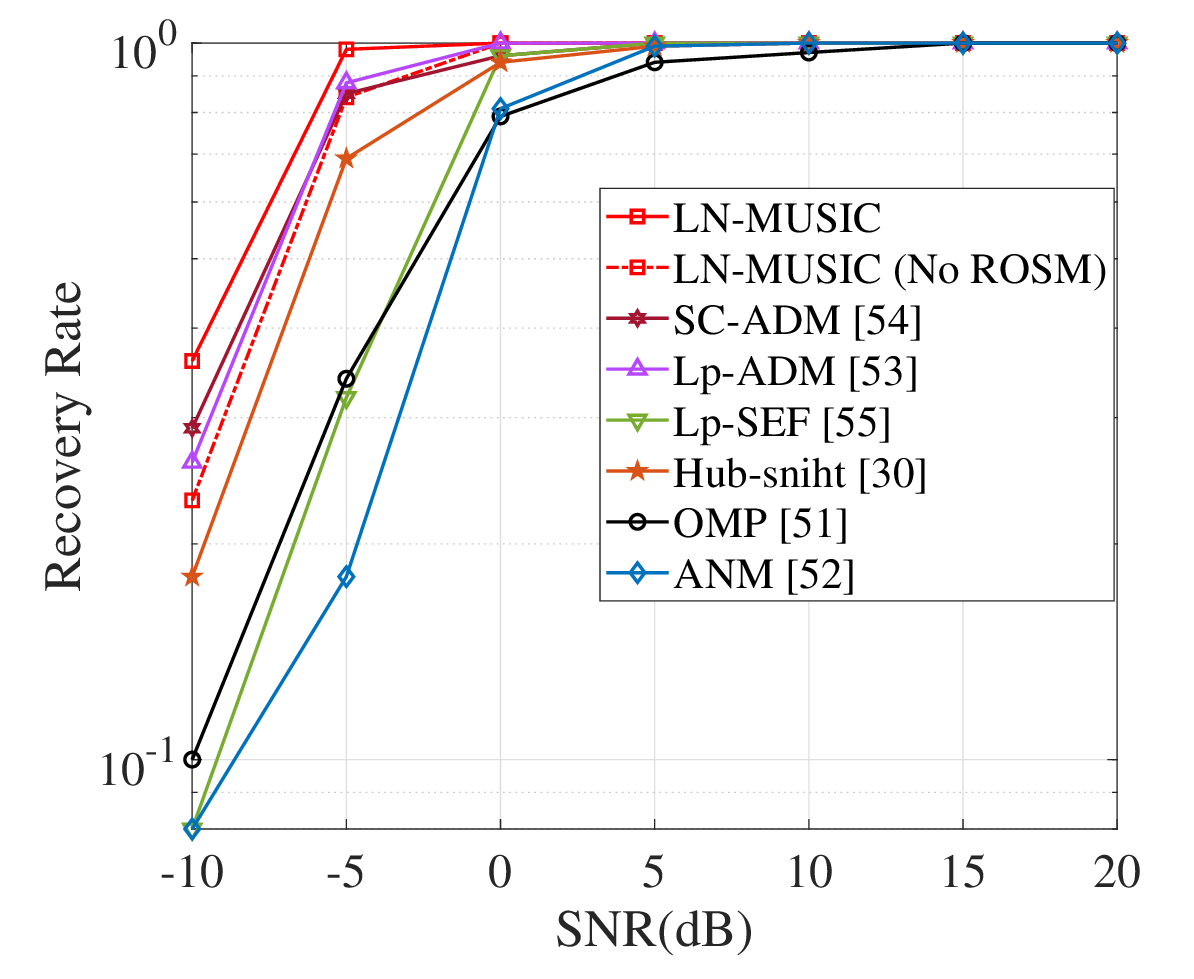}%
    \caption{Comparison of Recovery Rate of Different Method.}
    \label{fig:rate}
    \end{figure}

Fig.\,\ref{fig:rate} compares the reliability of several algorithms, which is measured by their recovery rate. The recovery rate of an algorithm is determined by performing 100 repetitions, dividing the estimated number of successful attempts by 100. For each algorithm, if RMSE$>5$ or cannot estimate $K$ angles, it is considered an estimation failure. The red line and the purple line in Fig.\,\ref{fig:rate} both represent {LN-MUSIC}, but the former makes use of the proposed ROSM strategy, while the latter does not. With an increasing of SNR, the recovery rate of all algorithms improves significantly. Among them, the reliability of {LN-MUSIC} is significantly better than other algorithms. When the SNR $= -5$ dB, almost every execution can ensure success.

For a DOA estimation algorithm, the computational complexity and estimation accuracy are the most important indicators to measure its performance. Especially in real-time systems or devices with limited computing power, the computational complexity is crucial. A lower computational complexity means that algorithms can run faster, reducing the demand for computing resources such as processors and memory. It is important for fast response, efficient resource utilization, and reduced power consumption. Table\,\ref{tab:time} shows the computational complexity of several existing DOA estimation algorithms and {LN-MUSIC}, wherein Hub-sniht is a grid-on DOA estimation algorithm and $grid$ denotes the number of spatial grids. {The computational complexity of LN-MUSIC is lower than that of the ANM and Hub-sniht, but higher than several other methods. We executed several methods on a personal computer equipped with a 12th Gen Intel(R) Core(TM) i7-12700 processor, obtaining the corresponding CPU run times for each method. With 100 Monte Carlo experiments, the CPU run time for LN-MUSIC is approximately 2 seconds, the fastest among all comparative methods is Lp-ADM, which requires about 1.2 seconds, while the slowest, the Hub-sniht method, exceeds 5 seconds.}
    \begin{table}[!htbp]
    \renewcommand\arraystretch{1.5}
    \centering
    \caption{Computational complexity} 
    \label{tab:time} 
    \begin{tabular}{ccc} 
    \toprule 
     \textbf{Method} & \textbf{Computational complexity} \\
    \midrule 
    OMP & ${\cal O}(MK^2)$\\
    ANM &${\cal O}( K ^{3.5})$ \\
    Hub-sniht&${\cal O}( grid^{2}K)$\\
    {Lp-ADM} &{${\cal O}( M^2K )$}\\
    {SC-ADM} & {${\cal O}(  M^2K )$}\\
    {LN-SEF} & {${\cal O}( MK^2 )$}\\
    \textbf{{LN-MUSIC}} & ${\cal O}( \left ( K + 1 \right )^{3})$\\
    \bottomrule 
    \end{tabular}
    \end{table}
    
\begin{figure}[htbp]
    \centering
    \includegraphics[width=3.5in]{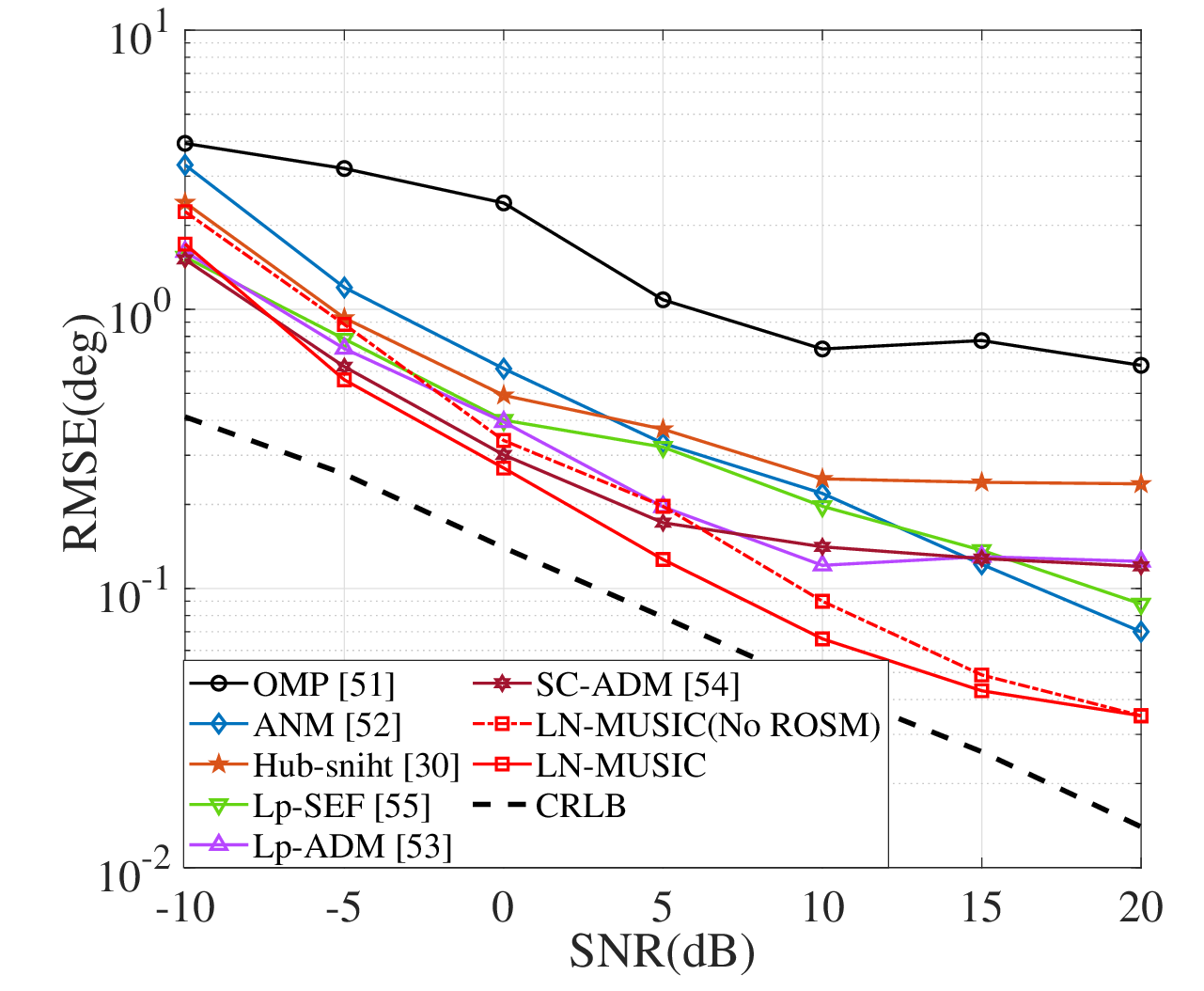}%
    \caption{Comparison of estimated performance under different noise.}
    \label{fig:compare}
    \end{figure}

Fig.\,\ref{fig:compare} demonstrates the estimation accuracy of LN-MUSIC compared to other algorithms across various SNR levels. It is evident that LN-MUSIC outperforms other algorithms, particularly for SNR ranging from $-10$ dB to 20 dB, where it demonstrates noticeably superior estimation accuracy. In particular, between 0 dB and 10 dB, the behavior of {LN-MUSIC} approach to the CRLB (the accuracy of DOA-estimation cannot fully approximate CRLB due to the consideration of path loss), which indicates that {LN-MUSIC} can effectively handle the interference of impulsive noises. In addition, considering the ROSM strategy proposed in this paper, from Fig.\,\ref{fig:compare} that the red line using the ROSM has a $30\%$ improvement in accuracy compared to the purple line without ROSM. This further validates the feasibility of the RIS control matrix optimization strategy proposed in this paper. 

    \section{Conclusions}
    
This paper has investigated the design of a RIS-based wireless communication system to achieve NLOS communication between targets and a single-antenna receiver, which simplifies the system structure. Based on the designed system, a robust DOA estimation algorithm, denoted LN-MUSIC, has been developed for scenarios with impulsive noises. In addition, this paper has proposed an RIS control matrix optimization strategy that does not require CSI, with a gain of up to $30\%$ for the DOA-estimation algorithm. Simulation experiments have demonstrated that {LN-MUSIC} achieves a superior performance to other popular methods. 

 \bibliographystyle{ieeetr}
\bibliography{RIS-LN}
\end{CJK}		
\end{document}